\documentclass[%
 reprint,
 superscriptaddress,
groupedaddress,
 nofootinbib,
 twocolumn, 
 amsmath,amssymb,
 aps,
 prl,
]{revtex4-2}
\usepackage{amsmath}
\usepackage{natbib}
\usepackage{amssymb}
\usepackage{xcolor}
\usepackage{graphicx}
\usepackage{dcolumn} 
\usepackage{bm} 
\usepackage[colorlinks = true,
            linkcolor = blue,
            urlcolor  = red,
            citecolor = red,
            anchorcolor = blue]{hyperref}
\usepackage{longtable}
\usepackage{gensymb}
\usepackage{siunitx}
\usepackage[version=4]{mhchem}
\usepackage{tabularx}
\usepackage[normalem]{ulem}
\usepackage{float}
\usepackage[caption=false]{subfig}
\DeclareUnicodeCharacter{2212}{-}

\newcommand{\nsno}{Nd$_{1-x}$Sr$_x$NiO$_2\,$}

\begin{document}
\texttt{}

\title{$\pi$/4 phase shift in the angular magnetoresistance of infinite layer nickelates}

\author{Yoav Mairovich}
\affiliation{Department of Physics, Ben-Gurion University of the Negev, Beer-Sheva, 84105, Israel}

\author{Ariel Matzliach}
\affiliation{Department of Physics, Ben-Gurion University of the Negev, Beer-Sheva, 84105, Israel}

\author{Idan S. Wallerstein}
\affiliation{Department of Physics, Ben-Gurion University of the Negev, Beer-Sheva, 84105, Israel}

\author{Himadri R. Dakua}
\affiliation{Department of Physics, Ben-Gurion University of the Negev, Beer-Sheva, 84105, Israel}

\author{Eran Maniv}
\affiliation{Department of Physics, Ben-Gurion University of the Negev, Beer-Sheva, 84105, Israel}

\author{Eytan Grosfeld}
\affiliation{Department of Physics, Ben-Gurion University of the Negev, Beer-Sheva, 84105, Israel}

\author{Muntaser Naamneh}
\thanks{Corresponding author: mnaamneh@bgu.ac.il}
\affiliation{Department of Physics, Ben-Gurion University of the Negev, Beer-Sheva, 84105, Israel}

\date{\today}

\begin{abstract}
The discovery of superconductivity in nickelates has generated significant interest in condensed matter physics. Nickelate superconductors, which are hole-doped within the layered structure of RNiO$_2$, share structural similarities with high-$T_c$ cuprate superconductors. However, despite similarities in formal valence and crystal symmetry, the fundamental nature of the superconducting state and the parent compound phase in nickelates remains elusive. Strong electronic correlations in infinite-layer nickelates suggest a potentially complex phase diagram, akin to that observed in cuprates, yet a key question about the magnetic ground state remains unanswered. Through magnetoresistance measurements across varying field strengths and orientations, we observe distinct angular-dependent magnetoresistance (AMR) oscillations with four-fold symmetry. Notably, this four-fold symmetry displays a $\pi/4$ phase shift with doping or applied magnetic field. Our findings parallel behaviors in electron-doped cuprates, suggesting that a static or quasi-static magnetic order exists in the infinite-layer nickelates, echoing characteristics of electron-doped cuprates. Furthermore, our modeling of the system reveals that the AMR is directly related to the underlying antiferromagnetic order, reinforcing this interpretation.
\end{abstract}


\maketitle


\emph{Introduction}.--- A central challenge in understanding unconventional superconductivity is unraveling how it emerges from adjacent phases whose interactions can either foster or inhibit superconductivity. This difficulty arises from the inherent complexity of these parent phases, exemplified by spin density wave order in Fe-based superconductors ~\cite{Stewart2011superconducticity} and charge density wave order in bismuthates ~\cite{Sleight2015}. Understanding these phases, along with the residual correlations that persist upon doping, is crucial for identifying the pairing mechanism in these materials. However, this remains a significant challenge, as the complex correlations within the parent phase, combined with additional interactions introduced by doping, often lead to competing or coexisting states and novel quasi-particles~\cite{Fradkin2015}.

Cuprates, renowned for their high-temperature superconductivity, are a prime example of this complexity~\cite{Keimer2015from}. Despite decades of intensive research, the mechanism by which superconductivity emerges from their insulating antiferromagnetic parent state remains elusive---particularly the nature of the pairing glue that binds electrons into Cooper pairs~\cite{anderson2007}.

\begin{figure*}[ht!]
\begin{center}
	\includegraphics[width=0.9\textwidth]{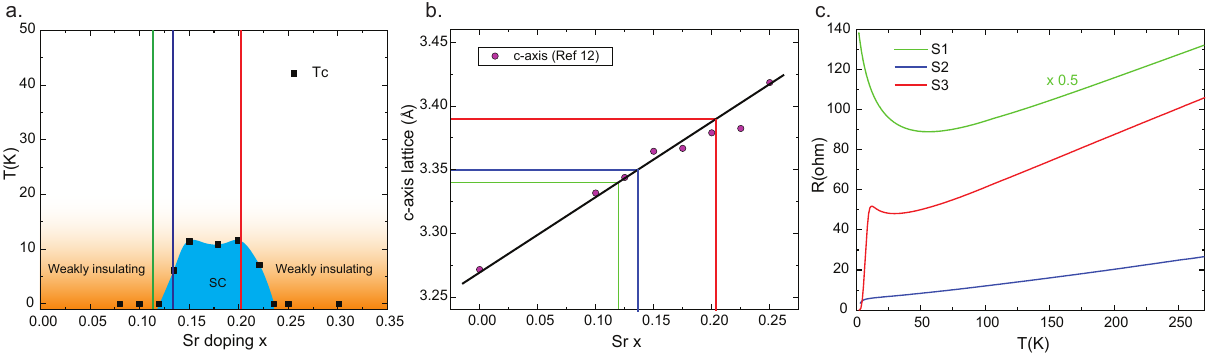} 
 \end{center}
	\caption{(a) Sketch of the $x$ - $T$ phase diagram of \nsno constructed according to Ref~\cite{Ariando2020}. The superconducting dome has a dome-like shape that peaks at $x=0.2$ corresponding to $3d^{8.8}$. The vertical lines mark the doping level of the three samples (S1-S3) of this study with $x=0.11,0.15,0.2$ respectively. (b) Room temperature c-axis lattice constant as a function of Sr substitution based on Ref~\cite{Hwang2020} The three lines mark the value of our three samples relative to the previous report. (c) Temperature-dependent resistance measured for the three samples. The resistance of sample S3 was divided by two for better visualization.}
	\label{fig1}
\end{figure*}

The recent discovery of superconductivity in infinite-layer (IL) nickelate thin films~\cite{li2019} has generated intensive interest, as IL-nickelates mirror key features of cuprates and may shed new light on their physics. One key similarity is the presence of NiO$_2$ planes where nickel adopts the same nominal electronic configuration $3d^9$ as copper in cuprates~\cite{pan2022}. Moreover, the superconducting state in IL-nickelates exhibits many parallels to cuprates, including a superconducting dome that peaks around $3d^{8.8}$ electron concentration upon hole doping [Fig.~\ref{fig1}(a)]. 

However, it is less clear whether the similarities persist in the parent compounds of these two families of materials. In cuprates, the undoped parent phase is an insulating antiferromagnet (AFM), with superconductivity emerging as AFM order is suppressed upon doping~\cite{Armitage}. In contrast, no definitive evidence for long-range AFM order has been observed in IL-nickelates, prompting ongoing efforts to determine their magnetic ground state and its dependence on doping and temperature~\cite{Sahinovic,Frank}.

Recent studies have provided signatures of AFM order in the undoped nickelate NdNiO$_2$. Resonant inelastic x-ray scattering has revealed magnon excitations similar to those observed in doped cuprates, albeit with a next-neighbor AFM exchange coupling only half that of cuprates~\cite{RIXS,Rossi}. Additionally, muon spin relaxation ($\mu$SR) measurements on superconducting Nd$_{0.8}$Sr$_{0.2}$NiO$_2$ suggest an intrinsic magnetic ground state, likely arising from local moments on the nickel sublattice, and indicate short-range AFM order~\cite{fowlie2022}. Despite these findings, the precise nature of magnetic order in IL-nickelates and its dependence on doping and temperature remains unclear. A key challenge is that, unlike cuprates, superconductivity in IL-nickelates has so far been observed only in thin films at ambient pressure. This precludes the use of traditional probes of magnetic order~\cite{Armitage,Kastner,musr2016}, which typically require bulk single crystals. Consequently, alternative methods are needed to shed light on the nature of their magnetic ground state.

In this paper, we investigate the dependence of the in-plane angular magnetoresistance (AMR) on hole doping in the IL nickelate \nsno. AMR measurements provide a sensitive probe of the spin structure in conducting samples via electron-spin interactions, a technique previously applied to cuprates, where it has been shown to detect spin reorientation, as confirmed by neutron scattering on bulk crystals~\cite{NCCO}. Our AMR measurements reveal a fourfold breaking of rotational symmetry in superconducting samples, consistent with previous reports~\cite{Wang2023,Ji2023}. Strikingly, this symmetry breaking persists in non-superconducting samples at lower doping but is rotated by $\pi/4$. Based on our calculations, we interpret the evolution of AMR symmetry with temperature and doping as a distinct signature of AFM ordering influencing conductance in both superconducting and normal states. Notably, the observed AMR symmetry closely resembles that of electron-doped cuprates, where it is also strongly tied to the AFM order of the parent compound~\cite{SLCO_AMR,NCCO,PLCCO}.

\emph{Growth and Characterization.}--- Three thin films of Nd$_{1-x}$Sr$_x$NiO$_3$, each with a thickness of $10\,$nm and labeled S1 to S3, were grown on a TiO$_2$-terminated (001) SrTiO$_3$ substrate using the pulsed laser deposition (PLD) technique with a $266\,$nm Nd:YAG laser. 
The c-axis lattice parameters determined based on X-ray diffraction (XRD) patterns of the three samples are shown in Fig.~\ref{fig1}(b). The analysis reveals that the c-axis lattice constant increases monotonically with Sr doping, following a linear trend consistent with a previous report of \nsno at room temperature~\cite{Hwang2020}.
Fig.~\ref{fig1}(c) shows the temperature-dependent in-plane resistivity for the samples. All samples (S1-S3) exhibit metallic behavior at high temperatures; however, sample S1 shows weakly insulating behavior at low temperatures, while samples (S2-S3) become superconducting with transition temperature $T_c$ that increases with Sr doping.

\begin{figure*}[ht!]
\centering
    \includegraphics[width=0.95\textwidth]{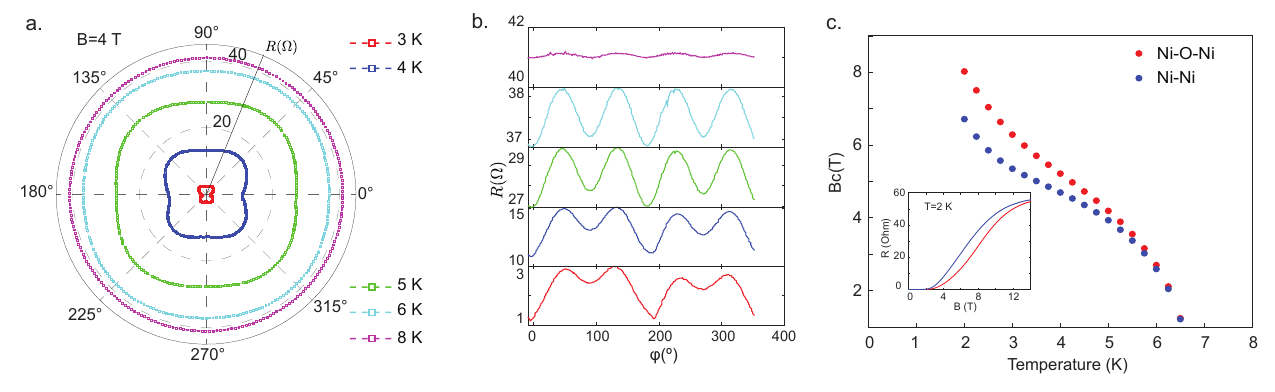} 
    \caption{In-plane angular magnetoresistance of superconducting \nsno sample (S3). The azimuthal angle dependence of the magnetoresistance at different temperatures under $B=4T$ is shown in (a) a polar plot and (b) a rectangular plot. (c) Temperature dependence of the critical field $B_c$ determined from magnetoresistance measurements under field along the Ni-O-Ni and Ni-Ni diriction. A representative magnetoresistance data obtained at temperature $T=2$K is shown in the inset of panel (c).}
	\label{fig2}
\end{figure*}

\emph{In-plane AMR}.--- In Fig.~\ref{fig2}, we present the in-plane AMR of the superconducting sample (S3), obtained by rotating the magnetic field in the plane while measuring the sample resistance. The resistance $R(\varphi)$ was measured through a fixed 4-probe contact configuration, with $\varphi$ representing the angle between the magnetic field and the a/b axis of the \nsno. Remarkably, the $R(\varphi)$ curves exhibit obvious four-fold rotational symmetry ($C_4$) at low temperatures as shown in the polar and rectangle plots in Fig.~\ref{fig2}(a-b). The $C_4$ symmetry is characterized by a minima at angles $0\degree,90\degree,180\degree,270\degree$, and maxima along angles $45\degree,135\degree,225\degree,315\degree$ relative to the a/b axis. This behavior is consistent with previous reports, and is independent of the measurement configuration~\cite{Wang2023,Ji2023}. The four-fold symmetry also includes a smaller two-fold symmetry $C_2$ component which can arise from a small tilt of the sample. However, the full symmetry-breaking pattern gradually restores isotropic symmetry as the temperature nears the superconducting transition [Fig.~\ref{fig2}(a-b)]. To gain more insight into the anisotropy of the AMR, we performed magnetoresistance measurements along the two highly symmetric directions, i.e., along Ni-O-Ni and Ni-Ni bonds corresponding to angles $0\degree$ and $45\degree$, respectively. From these measurements we extracted the superconducting critical field $B_c$, defining it as the field at half of the superconducting transition. Remarkably, the critical field was found to be anisotropic and correlated with the $C_4$ symmetry of the AMR, signaling their close relation [Fig.~\ref{fig2}(c)]. We should note that at temperatures adjacent to the transition temperature, both the critical field and the AMR are isotropic, while at lower temperatures, they similarly break the rotational symmetry to $C_4$. This could suggest a coexistence of superconductivity with another order which is responsible for the AMR symmetry breaking.

\begin{figure*}[ht!]
\centering
    \includegraphics[width=0.95\textwidth]{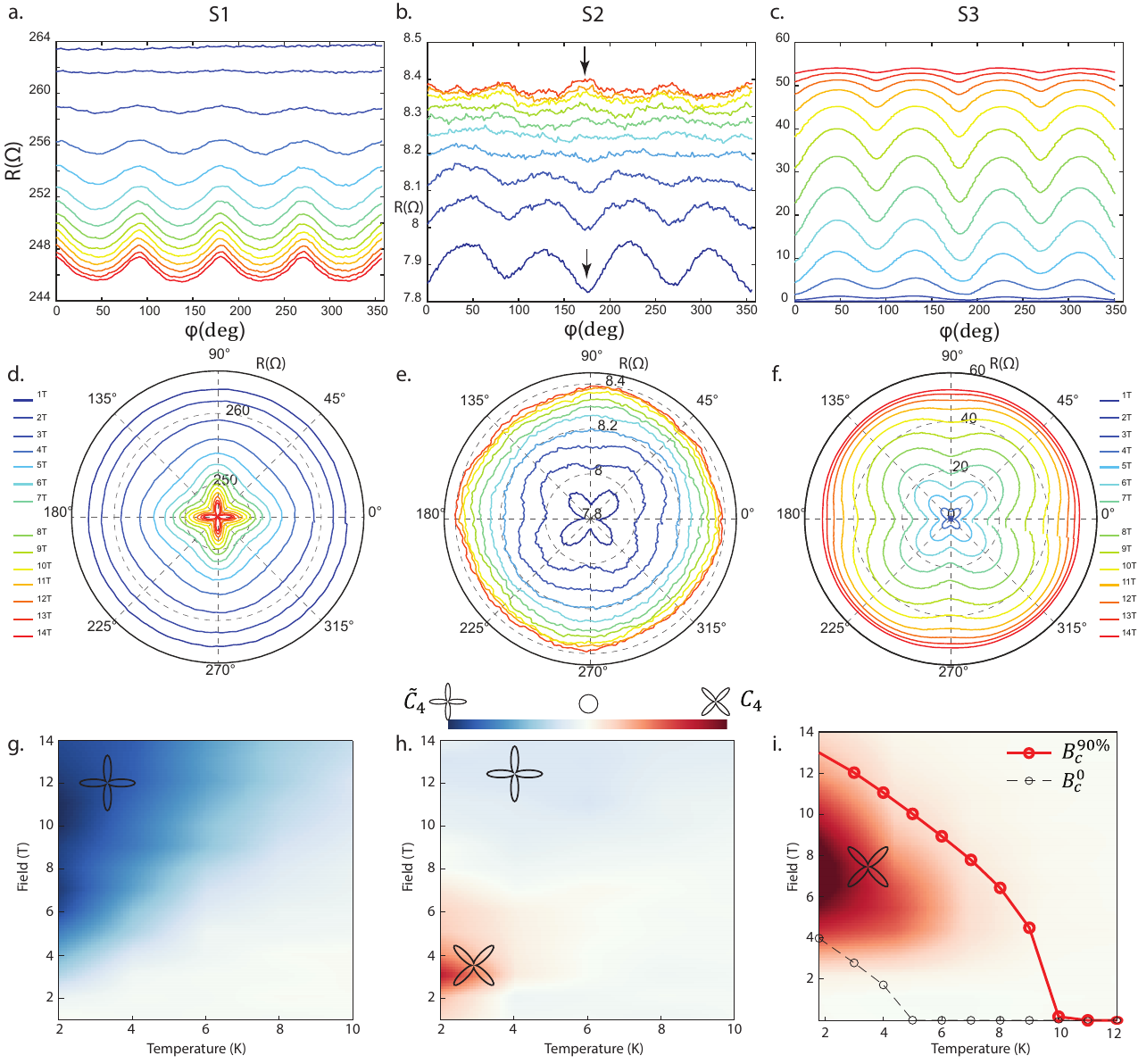} 
    \caption{The B-T phase diagram of \nsno samples (S1–S3) illustrates the angular magnetoresistance (AMR) symmetry. The azimuthal angle dependence of magnetoresistance at $T=2K$ under different magnetic fields is presented in both rectangular (a–c) and polar (d–f) plots. The results reveal that the AMR exhibits a fourfold symmetry, transitioning from $C_4$ to $\tilde{C}_4$ depending on doping, temperature, and magnetic field. The arrow in panel (b) highlights this transition, where the minimum resistance evolves into maximum resistance with increasing magnetic field. Panels (g–i) display the B-T phase diagram of the AMR magnitude, defined as $\delta R_{AMR}=R(\phi=45)-R(\phi=0)$. The $C_4$  symmetry corresponds to a positive $\delta R_{AMR}$, represented in red, while the $\tilde{C}_4$  symmetry corresponds to a negative $\delta R_{AMR}$, shown in blue.}
	\label{fig3}
\end{figure*}

When a magnetic field is applied to a superconductor, two primary mechanisms can break Cooper pairs: the orbital limiting effect and the Pauli paramagnetic effect. The orbital limit arises when the kinetic energy associated with supercurrents exceeds the superconducting gap, disrupting the pairing. In contrast, the Pauli limit is set by the Zeeman energy of the electronic spins, which, when exceeding the superconducting gap energy, renders singlet Cooper pairs energetically unstable. In the case of IL nickelates, superconductivity is believed to be two-dimensional, confined within the NiO$_2$ planes, which are oriented perpendicular to the $c$-axis~\cite{sun2023evidence,xiao2024superconductivity,chow2023dimensionality}. Consequently, the orbital limiting effect is expected to dominate when the magnetic field is applied perpendicular to the NiO$_2$ planes, while the Pauli effect plays a more significant role when the field is parallel to the planes. 
Recent experimental observations suggest a violation of the Pauli limit for in-plane magnetic fields~\cite{wang2021isotropic,sun2023evidence}.
However, if the Pauli limit were the sole determining factor, $B_c$ would be independent of the in-plane angle of the applied field. The AMR we observe thus indicates the presence of additional mechanisms that break the expected symmetry, such as magnetic ordering or another form of electronic ordering that influences the superconducting state.   



To investigate the origin of the symmetry breaking and its connection to superconductivity, we measured the AMR of a weakly insulating, lower-doped sample (S1). As shown in Fig.~\ref{fig3}(a), the results reveal two key differences from the superconducting sample. First, the AMR exhibits a fourfold symmetry, denoted as $\tilde{C}_4$, which is rotated by $\pi/4$ relative to the $C_4$ symmetry observed in superconducting samples. Consequently, $\tilde{C}_4$ has minima at $45\degree, 135\degree, 225\degree, 315\degree$ and maxima at $0\degree, 90\degree, 180\degree, 270\degree$ relative to the a/b axis. Second, sample S1 exhibits negative magnetoresistance, in contrast to the positive magnetoresistance observed in the superconducting sample.

Interestingly, while the insulating sample S1 maintains $\tilde{C}_4$ symmetry across all fields up to 14~T, the underdoped superconducting sample S2 undergoes a field-dependent transition, shifting from $C_4$ symmetry at low fields to $\tilde{C}_4$ symmetry at high fields. This suggests a common origin for the AMR symmetry breaking, manifesting as $C_4$ in the superconducting state and $\tilde{C}_4$ in the normal state.

Fig.~\ref{fig3}(g-i) summarizes the variation of the in-plane AMR across the \nsno phase diagram. In the insulating sample S1, the AMR exhibits $\tilde{C}_4$ symmetry that becomes more pronounced with increasing field. As doping increases, superconductivity emerges, accompanied by a progressively higher $T_c$. In the underdoped superconducting sample S2, the AMR initially displays a $C_4$ symmetry at low fields, persisting up to temperatures near $T_c$. Notably, at higher fields beyond the critical field, the symmetry transitions to $\tilde{C}_4$. Finally, the optimally doped sample S3 retains a $\tilde{C}_4$ symmetry at finite temperatures and fields that are associated with the superconducting transition, but ultimately becomes isotropic in the normal state. As we now argue, this behavior suggests parallels with the AFM correlations in electron-doped cuprates, where magnetic ordering persists within the superconducting state.

\begin{figure*}
    \centering
    \includegraphics[width=0.99\linewidth]{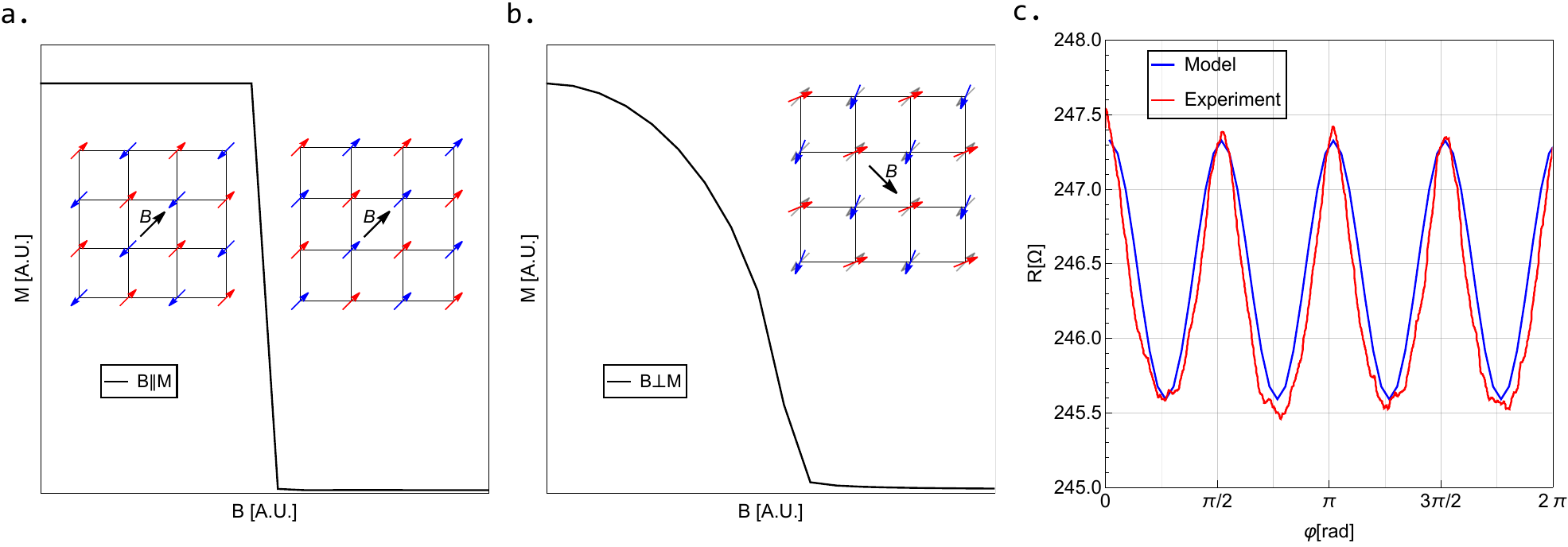}
    \caption{(a) First order phase transition for the staggered magnetization $M$ as function of $B\parallel M$; (b) Second order phase transition for the staggered magnetization $M$ as function of $B\perp M$; (c) AMR - experiment vs theory.}
    \label{fig4}
\end{figure*}

The results generally align with a physical scenario where the underlying AFM's staggered magnetization alternates its orientation between \( +45\degree \) and \( -45\degree \) directions within, e.g., two consecutive layers. Applying an external magnetic field along the staggered magnetization will generate a first-order phase transition where the magnetization abruptly goes to zero at a critical magnetic field $B_c$ [Fig.~\ref{fig4}(a)]. Instead, when the magnetic field is perpendicular to the staggered magnetization, it undergoes a softer, second-order transition to zero [Fig.~\ref{fig4}(b)]. Assuming an activated behaviour in the gap that is generated by the staggered magnetization, this will lead to a $\tilde{C}_2$-type AMR in each layer, rotated with respect to each other by $\pi/2$. When the conductances of two layers are added in parallel, the resulting resistance will carry a $\tilde{C}_4$-type symmetry as observed in the experiment. This framework also favors negative magnetoresistance~\cite{author2025supplement}.

In contrast, in the superconducting (SC) phase, a magnetic field induces a net magnetization that is most pronounced when the field is perpendicular to the staggered magnetization. This net magnetization suppresses the SC order parameter, thereby reducing the critical field strength and increasing resistance. For the observed $C_4$ symmetry, again two layers are required. 

This mechanism underpins the observed \(\pi/4\)-rotation in the AMR between the insulating and superconducting phases. More detailed calculations are provided in \cite{author2025supplement}, and a quantitative comparison with experimental data is shown in Fig.~\ref{fig4}(c).


\emph{Conclusions}.--- Superconducting IL nickelates exhibit AMR with four-fold symmetry, a behavior previously attributed to either possible d-wave pairing symmetry or the magnetic moment of the rare-earth element. These interpretations suggest distinct mechanisms with unique implications. If the four-fold symmetry were tied to d-wave pairing, one would expect it to disappear when superconductivity is suppressed. Consistent with this hypothesis, our results show that AMR becomes isotropic when superconductivity is suppressed by either temperature or magnetic field.

However, measurements on the underdoped and insulating samples reveal that the four-fold symmetry persists beyond the superconducting state, challenging the idea that it originates solely from pairing symmetry. 

An alternative explanation attributes the AMR to the magnetic moments of Nd atoms in the spacer layers between the Ni-O planes. This view is supported by the absence of four-fold symmetry in other rare-earth-based IL nickelates, such as La$_{1-x}$Sr$_x$NiO$_2$ and Pr$_{1-x}$Sr$_x$NiO$_2$~\cite{Wang2023}. If this were the case, one would expect the four-fold symmetry to remain robust against small variations in doping within the Ni-O planes.

Contrary to this expectation, our results demonstrate a dependence of the AMR on both doping and temperature. This strongly suggests that AMR is closely linked to an ordered state that evolves with these parameters and may even coexist with superconductivity. Comparing these findings with electron-doped cuprates, we observe striking similarities in the evolution of AMR, which in those systems is associated with AFM ordering, as informed by neutron scattering.

\emph{Acknowledgements} This research was supported by the ISRAEL SCIENCE FOUNDATION (grant No.~2509/20).

\bibliography{refs.bib}
\end{document}


\texttt{}

\title{Supplementary information for `$\pi$/4 phase shift in the angular magnetoresistance of infinite layer nickelates'}

\author{Yoav Mairovich}
\affiliation{Department of Physics, Ben-Gurion University of the Negev, Beer-Sheva, 84105, Israel}

\author{Ariel Matzliach}
\affiliation{Department of Physics, Ben-Gurion University of the Negev, Beer-Sheva, 84105, Israel}

\author{Idan S. Wallerstein}
\affiliation{Department of Physics, Ben-Gurion University of the Negev, Beer-Sheva, 84105, Israel}

\author{Himadri R. Dakua}
\affiliation{Department of Physics, Ben-Gurion University of the Negev, Beer-Sheva, 84105, Israel}

\author{Eran Maniv}
\affiliation{Department of Physics, Ben-Gurion University of the Negev, Beer-Sheva, 84105, Israel}

\author{Eytan Grosfeld}
\affiliation{Department of Physics, Ben-Gurion University of the Negev, Beer-Sheva, 84105, Israel}

\author{Muntaser Naamneh}
\thanks{Corresponding author: mnaamneh@bgu.ac.il}
\affiliation{Department of Physics, Ben-Gurion University of the Negev, Beer-Sheva, 84105, Israel}

\date{\today}


\maketitle

\section{Growth} 
\begin{figure*}
    \centering
    \includegraphics[width=0.40\linewidth]{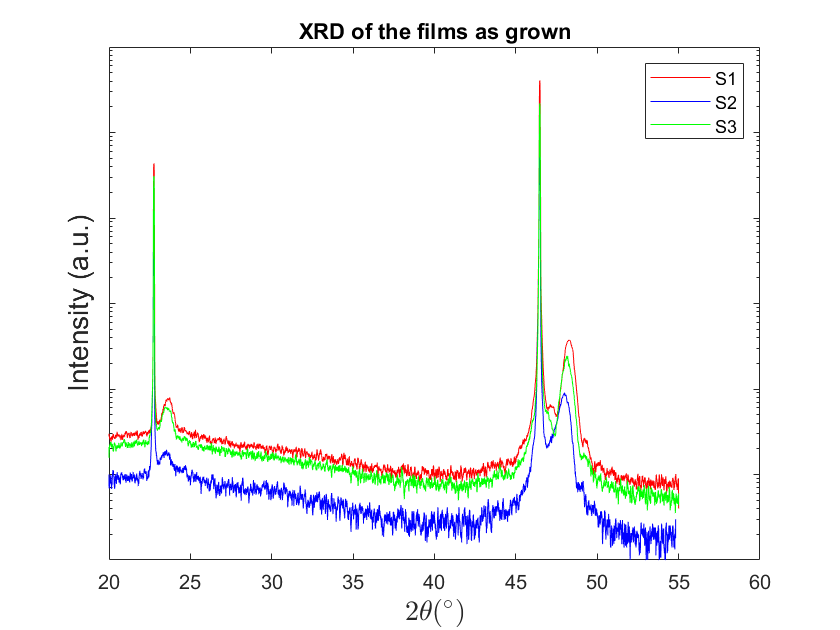}
    \includegraphics[width=0.40\linewidth]{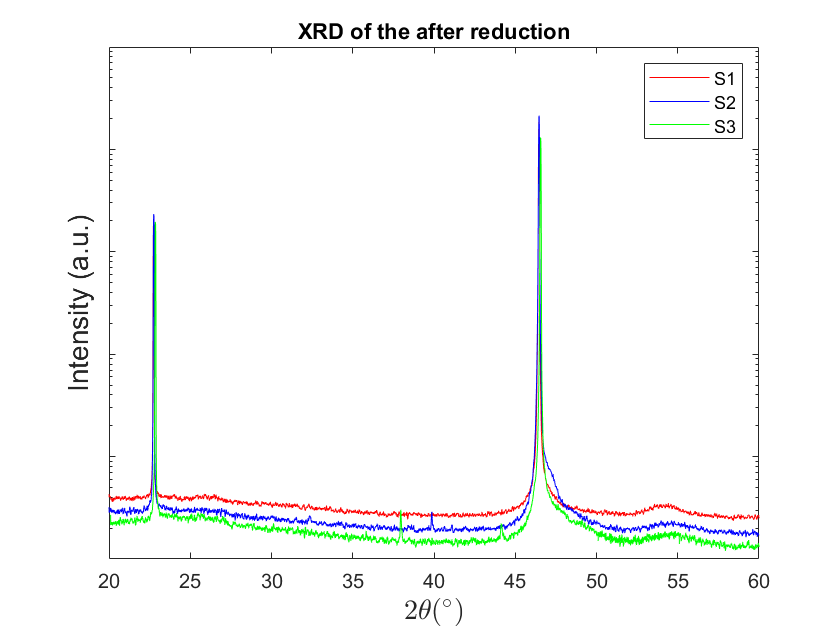}
    \caption{X-ray diffraction pattern of the three sample after reduction}
    \label{XRD}
\end{figure*}

Three thin films of Nd$_{1-x}$Sr$_x$NiO$_3$, each with a thickness of 10 nm and labeled S1 to S3, were grown on a $TiO_2$-terminated (001) SrTiO$_3$ substrate using the pulsed laser deposition (PLD) technique with a 266 nm Nd:YAG laser. The nominal composition of the as-grown films is assumed to match that of the target. No capping layers were applied to any of the samples in this study. The deposition process was conducted at a temperature of 630 °C and an oxygen partial pressure ($P_{O_2}$) of 150 mTorr for all samples. After the deposition, the sample were cooled down in the same pressure at the rate of 10$^\circ$/min. To obtain the infinite layer phase, the apical oxygen was removed using a soft-chemistry topotactic reduction method. The samples were sealed in quartz tubes in about 10$^-4$ Torr with 0.15g CaH$_2$ powder. Then, the tube is heated up to 280 $^\circ$C for 2 h. This process induced a topotactic phase transition in the nickelate thin films, converting them from the perovskite phase to the infinite-layer phase (\nsno) as shown in Figure.~\ref{XRD}.

\section{Transport}  The samples were contacted using wire-bonded aluminium wires in a 4-probe configuration. The angle-dependent transport measurements were carried out on a rotator with an accuracy of 0.01$^{\circ}$ in a 14 T-Physical Property Measurement System (PPMS-EverCool-14, Quantum Design).

\section{Model} 
Following \cite{Fradkin_2013}, we consider the Hubbard model at half-filling with the addition of the external in-plane magnetic field $\mathbf{B}=B\cos\varphi\hat{x}+B\sin\varphi\hat{y}$
\[H =  - t\sum\limits_{\left\langle {i,j} \right\rangle,\sigma }^{} {c_{\sigma ,i}^\dagger {c_{\sigma ,j}}}  + \sum\limits_i^{} {\left( {{\mathbf{B}} - \frac{{2U}}{3}{{\mathbf{S}}_i}} \right) \cdot {{\mathbf{S}}_i}} \]
where $t$ is the hopping amplitude, $U$ is the on-site interaction.
The first summation is over all nearest neighbor bonds and spin $\sigma\in\{\uparrow,\downarrow\}$, while the second summation is over all sites.
The operators $c_{i,\sigma}(c^\dagger_{i,\sigma})$ are the annihilation (creation) of a $\sigma$-spin electron at site $i$, and
\[{{\mathbf{S}}_i} = \frac{1}{2}\sum\limits_{\rho,\tau}c_{\tau ,i}^\dagger {\bm\sigma _{\tau \rho }}{c_{\rho ,i}}\]
where $\bm\sigma$ is the Pauli matrices vector.

Introducing the magnetization ${\bf M}_i=-(4U/3)\langle{\bf S}_i\rangle$, where ${\bf M}_i$ is staggered between neighbor lattice sites, and using the Hartree-Fock mean-field approximation, we arrive to the one-particle $k$-space mean-field Hamiltonian (sans a constant term)
\[{\mathcal{H}_k} = \left( {\begin{array}{*{20}{c}}
  { - {t_k}}&{\frac{1}{2}\left( {{\mathbf{M}} + {\mathbf{B}}} \right) \cdot \bm{\sigma} } \\ 
  {\frac{1}{2}\left( {{\mathbf{M}} + {\mathbf{B}}} \right) \cdot \bm{\sigma} }&{{t_k}} 
\end{array}} \right)\]
where $t_k=2t(\cos{k_x}+\cos{k_y})$ and the Brillouin zone is folded, i.e., $-\pi\le k_x\le\pi,0\le k_y\le\pi$.
The eigenenergies are
\[ \pm {E_ {\pm,k}} =  \pm \frac{1}{2}\sqrt {{B^2} + {M^2} + {{\left( {2{t_k}} \right)}^2} \pm 2\sqrt {{{\left( {{\mathbf{B}} \cdot {\mathbf{M}}} \right)}^2} + {B^2{\left( {2{t_k}} \right)}^2}} } \]
At half-filling, in zero temperature, the two negative energy bands are filled.
In our calculations, we assume that the staggered magnetization is aligned to a fixed orientation $\theta$, such that the resulting magnetization has an angular dependency of $\cos^2(\varphi-\theta)$.
The mean-field equation is then ${\partial _M}\mathcal{E}=0$, where $\mathcal{E}$ is the energy functional
\[
{\mathcal{E}}=\frac{3M^2}{8U}-\frac{2}{N_xN_y}\sum\limits_{\begin{gathered}
  k_i=2\pi n_i/N_i\\
  0\le n_i<\lfloor N_i/2\rfloor\\\ 
\end{gathered}}{(E_{+,k}+E_{-,k})}
\]
where $N_i$ is the number of sites in the $i$ direction.
We take the numerically convergent resolution of $N_x=N_y=30$. 

\begin{figure*}
    \centering
    \includegraphics[width=0.30\linewidth]{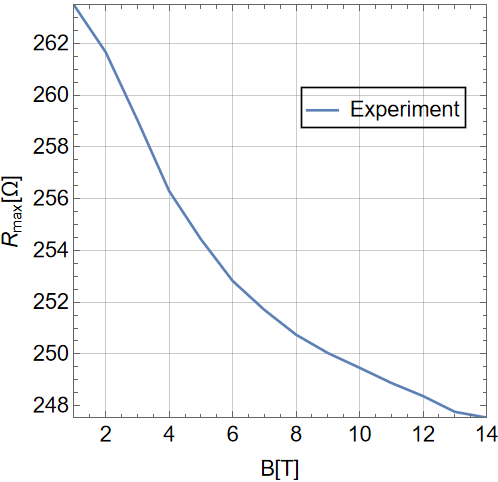}
    \includegraphics[width=0.30\linewidth]{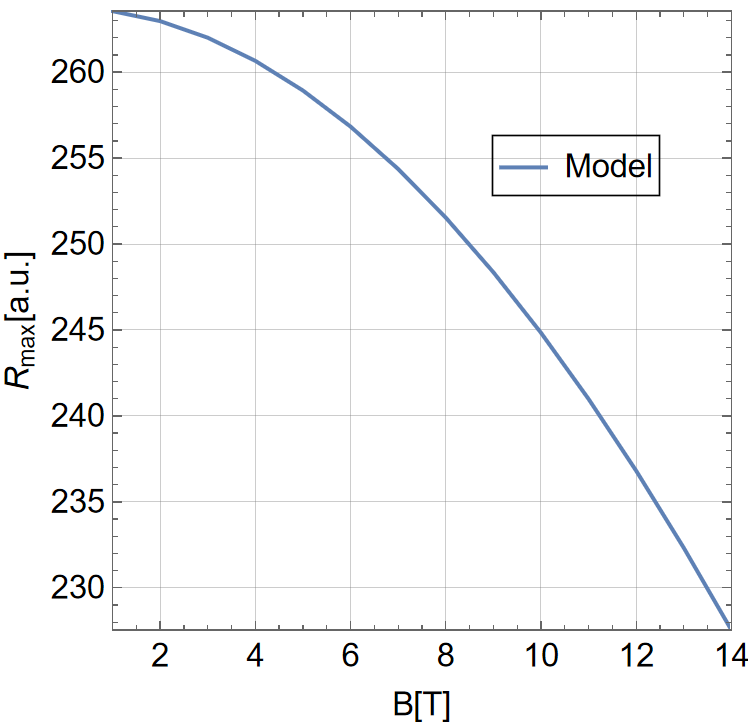}
    \caption{Negative magnetoresistance. (a) Experimental; (b) Model (given in arbitrary units).}
    \label{fig_sup}
\end{figure*}

Fig.~\ref{fig1}(c) suggests an activated behavior, hence, for simplicity, we assume that the conductivity is given by 
\[\sigma(\varphi,\theta)=\sigma_0\exp[-M(\varphi-\theta)/k_BT]\]
or, equivalently
\[R(\varphi,\theta)=R_0\exp[M(\varphi-\theta)/k_BT]\]
where the magnetization is approximately the gap.
Considering two layers, each with a fixed orientation of $\theta=\pm\pi/4$, the resistance is
\[R(\varphi)^{-1}=R(\varphi,+\pi/4)^{-1}+R(\varphi,-\pi/4)^{-1}\]
For Fig.~\ref{fig5}(b), the Hubbard model parameters are taken as $t=0.2eV$ and $U=0.115eV$.
The free parameter $R_0$ is chosen to give the correct scale for the resistance at $B=14T$.
In Fig.~\ref{fig_sup}, we plot for the model and the experiment the magnetoresistance, i.e., the variation of the maximum resistance, $R_{\max}=R(\varphi=0)$, with strength of the magnetic field.
b
\pagenumbering{gobble}

\bibliography{refs.bib}